\documentstyle[12pt,aps]{revtex}
\begin{document}

\title{\vspace{0.5cm}
         Stability Properties of  $|{\Psi}|^2$ in Bohmian Dynamics}

\author{G. Potel, M. Mu\~{n}oz-Ale\~{n}ar,}

\address{
     \vspace{.2cm}
     {\setlength{\baselineskip}{18pt}
     Dpto. de F\'{\i}sica Aplicada I, Universidad de Sevilla, Spain.}}

\author{F. Barranco}

\address{
     \vspace{.2cm}
     {\setlength{\baselineskip}{18pt}
     Dpto. de F\'{\i}sica Aplicada III, Universidad de Sevilla, Spain.}}

\author{and}

\author{E. Vigezzi}

\address{
     \vspace{.2cm}
     {\setlength{\baselineskip}{18pt}
     INFN, Sez. Milano, via Celoria 16, 20133 Milano, Italy.}}
\vspace{.5cm}
\maketitle
\pagestyle{empty}
\begin{abstract}

 According to Bohmian dynamics, the
 particles of a quantum
system move along trajectories, following a velocity field
determined by the wave-function $\Psi(x,t)$. We show that
for simple one-dimensional systems any initial probability
distribution of a statistical ensemble approaches
asymptotically $|{\Psi(x,t)}|^2$ if the system  is subject
to a random noise of arbitrarily small intensity.
\end{abstract}

%\pacs{{\sl Subject headings}: Bohmian dynamics --
%probability distributions\\

\vspace{5mm}


%\newpage

\vspace{2cm}

\pagestyle{plain}

According to the de Broglie-Bohm  quantum theory of motion
[1-3], the $n$ particles of a  quantum system described by
a wavefunction
\begin{equation}
 \Psi(\vec X,t) = R(\vec X ,t )  e^{[(i/\hbar)S(\vec
 X,t)]}, \quad \vec X \equiv\{{\vec x_1},{\vec x_2},...{\vec x_i},...,{\vec x_n}\}
% \in {\mathcal R}^{3n},
\label{1}
\end{equation}
that obeys the Schr\"{o}dinger equation, follow trajectories
whose velocity is given by the equation
\begin{equation}
\dot{\vec x_i} = \frac{1}{m} \frac{ \partial S(\{{\vec x_1},{\vec x_2},...{\vec x_i},...,
{\vec x_n}\}, t)
}{\partial{\vec x_i}} |_{\vec X = \vec X(t)}, \quad
i=1,...,n
\label{2}
\end{equation}
which  can be solved knowing the initial conditions $ \vec
X(t=0)$. Note that in this equation the actual value of the
velocity (in configuration space) is the realization of a
particular value of the velocity field $ 1/m \nabla_{\vec
X} S(\vec X , t)$ at the actual position $\vec X(t)$ of the
system.

Consider now an ensemble of systems, associated with a
probability distribution in configuration space, $\rho(
\vec X, t)$. Both $\rho$ and the quantum distribution
probability, $| \Psi |^2$ , obey a continuity equation [2],
\begin{equation}
\partial g(\vec X,t) / \partial t + \nabla_X \cdot ( g(\vec X,t)  \dot{\vec X}) =
0, \quad g=\rho, |\Psi|^2.
\label{2b}
\end{equation}
It is clear that if $\rho$ and $ | \Psi |^2$ coincide at a
given time $t_o$, they coincide at all times (that is
called the equivariance property) so that the predictions
of quantum mechanics are reproduced. The condition
$\rho(\vec X, t_o) = | \Psi (\vec X, t_o) |^2$ is then
sometimes taken as a postulate of the Bohmian quantum
theory of motion (cf. e.g. [3]). However, this restriction
of the theory to particular ensembles is not logically
justified, since in the theory trajectories have an
independent reality of their own [4], and in fact there
have been attempts to deduce the quantal distribution from
the dynamics of the system. Precisely because of the
existence of trajectories this problem can be considered to
be akin to that of classical statistical mechanics [5].
Valentini [6,7] has deduced a "subquantal" H-theorem, which
involves hypothesis similar to those adopted in the
demonstration of the H-theorem in classical mechanics,
assuming that the orbits of the system are sufficiently
complicated, so that they sample all the accessible regions
of configuration space. Indeed, many recent studies deal
with the chaotic features of the deBroglie and Bohm
trajectories for systems with several degrees of freedom
(cf. e.g. [8]).

We shall follow a different line of thought, inspired by the
original paper by Bohm [9], who studied the consequences
of random collisions among the electrons inside atoms or molecules,
on the electron distribution.

Let us suppose that the electrons of a given statistical
sub-ensemble of molecules are initially described by the
stationary wavefunction of the ground state,
$\psi_a(x_1,t)$ , and by a distribution of particles, not
necessarily corresponding to $|\psi_a|^2$. Let us further
suppose that, during some amount of time, these molecules
interact with a second set of uncontrolled molecules,
described by the wavefunction $\phi_b(x_2,t)$.

Initially (for $t\rightarrow-\infty$) the wavefunction of the two colliding
molecules is
\begin{equation}
\Psi_o(x_1,x_2,{\mathbf{r}}_{12};t)= \psi_a(x_1;t) \phi_b(x_2;t)
d_o({\mathbf{r}}_{12};t),
\label{9}
\end{equation}
while during the collision
the wave function must be written as the linear combination
\begin{equation}
\Psi(x_1,x_2,{\mathbf{r}}_{12};t) = \sum_{\alpha,\beta,\gamma}
{\mathcal{C}}_{\alpha,\beta,\gamma}
\Psi_{\alpha,\beta,\gamma}(x_1;\{ x_2,\mathbf{r}_{12};t\})
\label{10}
\end{equation}
with

\begin{equation}
\Psi_{\alpha,\beta,\gamma}(x_1,x_2,{\mathbf{r}}_{12};t)= \psi_\alpha(x_1;t)
\phi_\beta(x_2;t)
d_\gamma({\mathbf{r}}_{12};t)
\label{11}
\end{equation}
where  $\psi_\alpha(x_1,t)$ represents a stationary state
of molecule-1, $\psi_\beta(x_2,t)$ represents a stationary
state of molecule-2, and $d_\gamma({\mathbf{r}}_{12},t)$
represents a  state of relative motion of the center of
mass of the colliding  molecules (typically a wave-packet).
The details of the initial state, that is $\psi_a(x_1),
\phi_b(x_2)$ and $d_o({\mathbf{r}}_{12})$, determine,
via the Schr\"odinger equation, the amplitudes
${\mathcal{C}}_{\alpha,\beta,\gamma}$ of the different
stationary states of the compound wave function.

Note that, in order to calculate the velocity of
electron-1, this wave function may be rewritten as the
linear combination of eigenstates of molecule-1

\begin{equation}
\Psi(x_1,t) = \sum_{\alpha}
{\mathcal{X}}_{\alpha}(x_2(t),{\mathbf r}_{12}(t);t)\psi_{\alpha}(x_1;t),
\label{12}
\end{equation}
in which the coefficients
${\mathcal{X}}_{\alpha}(x_2(t),{\mathbf{r}}_{12}(t);t)$
incorporate the dependence in $x_2$ and
${\mathbf{r}}_{12}$, and the summation over the indices
$\beta$ and $\gamma$. The velocity of the electron in the
interior of molecule-1 is obtained using this wave function
by means of Eq.(2), being thus sensitive to all the initial
conditions, including, the
positions of the electrons of molecule-2, ($x_2(t=-\infty)$)
and the relative positions.
Similar expressions may be adopted for the
velocities associated to $x_2$ and ${\mathbf{r}}_{12}$.

After the collision (for $t\rightarrow+\infty$) the linear
combination in Eq.(5) may be simplified, for all practical
purposes, by retaining only one component, for example that
of $\alpha=\beta=a$ in the case of elastic scattering. This
is a kind of "natural" wave function collapse (see [2,10]),
and guarantees that after the collision process the system
is left in a stationary state. The question one would like
to address is how the collision process affects the
original distribution.

In order to find an answer, we shall study the evolution of
the particle distribution of one of the subensembles, that
of coordinate $x_1$, replacing the detailed description of
the other variables ($x_2$ and ${\mathbf{r}}_{12}$) during
the collision process by the action of a random noise. The
stochastic noise will represent the effect of the deviation
of all those other degrees of freedom from their average value.

We shall
solve the equations of motion, studying numerically the
resulting distribution function of the electrons, instead of
making the
strong statistical assumptions adopted by Bohm
\footnote{Bohm's proof was based on the assumption
of a continuity equation (Eq.(3) in [9]), even if the
random perturbation acting on the electrons inside the
molecules would require the more general Fokker-Planck
equation, including sources and diffusion terms.}. Our aim
will be to show that the system tends to a distribution
described by $|\Psi|^2$, starting from an arbitrary initial
distribution.

We first study some properties of the time evolution of the
distribution of particles governed by Bohmian dynamics, in
relation with its response to a stochastic perturbation. In
particular we consider the effects of stochastic noise on
one-dimensional, non-degenerate systems in non-stationary
states. Starting from arbitrary initial ensemble
distributions,  we will show that they tend to the quantal
distribution.

In a  realistic calculation,
the noise intensity should
be determined in accordance to the coefficients of the
linear combination (7), and as a  consequence should be dependent
on the position along the trajectory.
It would be  natural to expect that during the
reaction the noise will reach a peak value around the
distance of closest approach between the colliding
molecules, tending to zero for
$t\rightarrow\pm\infty$.
In the following, we shall limit ourselves to a very simplified description,
solving the equations of
motion (the one-dimensional equivalent of Eq.(2))
\begin{equation}
\dot{x}_i = 1/m \nabla S(x_i,t) + C \eta_i,
\quad i = 1,2,...N
\label{3}
\end{equation}
where $i$ labels the $N$ elements of the statistical
ensemble of one-particle systems, initially distributed
according to a given function $\rho_0(x) \equiv
\rho(x,t=0)$, and subject to a random white noise $\eta_i
(t)$, whose intensity is controlled by the parameter $C$
(in what follows $C$ will be measured in hundredths of the
speed of light, which is a natural scale for the studied
systems).
The $x$-independent character of the noise
that we use, would produce a complete diffusion in the
whole space. This is avoided by confining the system in a box
with infinite, perfectly reflecting
walls.

Let us thus numerically study the case of a statistical
ensemble of $N= 5000$ test systems, each of them
representing an electron confined in a box of length $L = 2
\AA$  with wave-function
\begin{equation}
\Psi(x,t) = a_1 \phi_1(x) e^{-iE_1t/\hbar} +
a_2\phi_2(x) e^{-iE_2t/\hbar},
\label{4}
\end{equation}
where $\phi_1$ and $\phi_2$ are the two first stationary
states of the well, taken with amplitudes  $a_1, a_2$,
which in a more general case would represent typical values
during the reaction process (cf. Eq.(7)).
Here for simplicity
we shall use the constant values $a_1 = a_2 = 1/\sqrt{2}$.
We have solved the stochastic differential equation (8)
using Heun algorithm [11].

We shall compare the results obtained with two different
initial distributions, a uniform one, $\rho_0(x)
=\rho_{un}(x)
\equiv 1/L$, and another coinciding, at $t=0$, with the
usual quantum distribution, $\rho_0(x)=
\rho_q(x) \equiv |\Psi(x,0)|^2$.
The time evolution of $\rho_0(x,t)$ starting from
$\rho_{un}(x)$ is shown in Fig.1 for the case of $C=0.1$,
which is  a rather large noise intensity, corresponding to
about $2\%$ of the average velocity . It is seen  that the
approach to $|\Psi (x,t) |^2 $ is already significant after
one oscillation of the distribution.

We shall measure the difference between the
distribution $\rho(x,t)$ evolving from the initial one,
$\rho_0(x)$, and the quantal distribution $|\Psi (x,t) |^2
$ making use of the function
\begin{equation}
\chi^2_{\rho_0}(t)
= \int{ |\rho(x,t)-|\Psi(x,t)|^2|dx}.
\label{5}
\end{equation}

    In Fig.2 we present the results
for three different values of $C$. For each one we present
the time evolution of $\chi^2$. To be noted that for both
initial conditions ($\rho_{un}(x)$ and $\rho_{q}(x))$ the
systems evolve under the influence of the random noise;
therefore, even starting from $\rho_q$, $\chi^2_{\rho_q}$
is in general different from zero. Similar results are
obtained using other measures of the difference between
$\rho(x,t)$ and $\Psi(x,t)$, in particular the relative
entropy used in [7,8].

It can be seen that $\chi^2_{\rho_{un}}$ decreases until it
reaches a minimum value, around which it stabilizes. This
value coincides (apart from statistical fluctuations
arising from the finite number of particles) with the one
attained by $\chi^2_{\rho_q}$ . This feature is seen in all
the three cases presented, although the time needed for
$\chi^2$ to stabilize is longer ($\tau$= 60, 600 and 6000 periods)
for smaller values of $C$ ($C$= 0.01,0.001 and 0.0001). It
is also important to note that the asymptotic value of
$\chi^2$ becomes smaller for smaller $C$, which means that
the obtained particle distribution is closer to the quantum
mechanical value $|\Psi(x,t)|^2$ the smaller the value of
$C$ is. This is illustrated by the distributions presented
on the right side of Fig.2.
%\footnote{
It can be seen that reducing the noise, $\chi^2$ tends to
an asymptotic value that is not equal to zero, what is
mainly ($90\%$) due to the finite statistical sampling
($5000$ systems distributed over $200$ x-subintervals),
and in a much smaller extent to the unavoidable errors in
the numerical integration of the equations of motion
(performed with $\Delta t=1/500$ of a period).
In fact, we have verified that
eliminating the noise, that is, putting $C$ equal to zero,
the asymptotic value is equal to $0.20$, very close to the
value obtained for $C=0.0001$.

In order to obtain some insight in the results discussed
above, it is convenient to rewrite the Bohmian equation of
motion as a second-order newtonian equation for the
acceleration[2],

\begin{equation}
m \ddot{x}_i = - \nabla U_{cl} - \nabla U_{qu},
%+ m Q d\eta_i/dt,
\label{6}
\end{equation}
where $U_{cl}$ is the classical potential and
\begin{equation}
U_{qu} = -(\hbar^2/2m)\nabla^2R/R
\label{7}
\end{equation}
is the so-called quantum potential.

 In the case of the particle in the box
$U_{cl}=0$ and thus, aside from the stochastic noise, the
trajectories are determined by the behavior of $U_{qu}$.
This is a time dependent potential, which goes to infinity
where $R=0$, that is at the nodal points. Instead at the
points where $R$ has maxima, $U_{qu}$ is almost flat. 

In the
absence of noise, the Bohmian trajectories may be thought
as those of classical particles moving in the time
dependent potential $U_{qu}$, with the condition that the
initial velocities, needed for integrating Eq.(11), must be
those determined by the wave-function of the system (cf.
Eq.(2)).

In our case, with only two eigenstates in the linear
combination of $\Psi$ (cf. Eq.(9)), the density oscillates
in time with Bohr's period $T = 2 \pi \hbar/(E_2-E_1)$, and
thus the potential $U_{qu}(x,t)$ and the trajectories are
also periodic (see Fig.3 and Figs. 4a,b) \footnote{In general, for
more complex linear combination of states of the box, the
periodicity will be controlled by the ground state
frequency, of which all the other
frequencies are multiples.}. The trajectories near to the nodal points
are compressed and re-expanded periodically by the strongly
varying quantum potential. Such a compression-expansion is
instead very soft for the trajectories close to maxima of
$R$ (note $R$ in the denominator in Eq.(12)).

The effect of the stochastic noise is to break this
periodic pattern (see Fig.4c). This breaking can be very
effective, because any slight perturbation in the
coordinate during the compressed phase makes the particle
dramatically change its subsequent trajectory in the
re-expansion process. The net effect of all these changes
of trajectory is a higher residence time around the maxima
of $R$, that produces a distribution of particles
resembling $|\Psi|^2$ in spite of the initial distribution.
We then see why the distribution $|\Psi|^2$ displays the
remarkable stability against random perturbations of the
system found above (cf. Fig.2).

 We have checked that the above results are
neither a special feature of the chosen potential, nor of
the particular considered wave-function, by repeating the
calculation with potentials of different shapes inside the
walls and with different linear combinations of
eigenstates, obtaining always a similar behaviour for the
density distribution.
 In the case of the parabolic  potential we have
also analyzed the classically forbidden region, that is the
region where the classical potential is larger than the
average energy of the considered wave function, and found
again that the density distribution approaches $|\Psi|^2$
as the noise intensity is set to smaller values. Let us
however make some specific comments on the case of
stationary states.

 According to Bohmian dynamics, in the case of a
stationary state ($\Psi(x,t) =\phi(x) e^{- i Et/\hbar}$)
the electrons are at rest (cf. Eq.(2)), and the addition of
a stochastic noise would give rise to a pure diffusion
process, leading inevitably to a uniform distribution
inside the box. However this situation is not physically
meaningful since, as stated above, the stochastic noise
represents the (random) deviations of the single-particle
like wave-function in eq.(7) from its average linear
combination, which for a pure stationary state is just zero.
In other words, the non-stationary character of the
wave-function and the stochastic noise go together.

We conclude that under the action of random noise, at least
for very simple, one-dimensional systems, an ensemble of
particles, initially arbitrarily distributed, and governed
by Bohmian dynamics progressively loses the "memory" of the
initial distribution as time proceeds, and approaches the
distribution $|\Psi(x,t)|^2$, a process which is based on
the different stability of the underlying Bohmian
trajectories. The origin of the random noise may be
naturally found in the uncontrollable character of the
position of the particles in Bohm's theory.

\newpage

{\bf Figure captions}

\vspace{5mm}

{\bf Fig. 1}

Evolution of the particles distribution during the first
oscillation corresponding to an initially uniform
distribution subject to a noise intensity of $C=0.1$
(continuous line) and the corresponding $|\Psi(x)|^2$
(dashed line).

\vspace{2mm}
{\bf Fig. 2}

 On the left hand side the evolution of
$\chi^2_{\rho_{un}}(t)$ (solid line) and
$\chi^2_{\rho_q}(t)$ (dashed line) is shown as a function
of time for three different values of $C$ (0.01, 0.001 and
0.0001). On the right hand side the distribution
$\rho(x,\tau)$ evolving from $\rho_{un} $ (solid line) for the
three different $C$ values at a fixed values of time ($\tau$=
60, 600 and 6,000 periods, respectively), is compared to
$|\Psi(x,\tau)|^2$ (dashed line).

\vspace{2mm}
{\bf Fig. 3 }

The periodic, time-dependent quantum potential $U_{qu}
(x,t)$ (in eV) is shown as a function of $x$ (in \AA) and
$t$ (in periods). Note the nodal point appearing as peaks
in $U_{qu}(x,t)$.

\vspace{2mm}
{\bf Fig. 4}

 a) Trajectories of a set of particles initially
distributed according to $|\Psi(x,t=0)|^2$, and without any
noise. b) Trajectories of a set of particles initially
uniformly distributed ($\rho_o =
\rho_{un} \equiv 1/L$) without any noise. c) The same as in b)
but under the action of the stochastic noise ($C= 0.1$ in this
example).

\end{document}